\pdfoutput=1
\documentclass[aps,prd,reprint,superscriptaddress,longbibliography,nofootinbib]{revtex4-1}
\usepackage{graphicx}
\usepackage{amsmath}
\usepackage{amssymb}
\usepackage{amsfonts}
\usepackage{dcolumn}
\usepackage{dsfont}
\usepackage{makecell}
\usepackage{latexsym}
\usepackage{rotating}
\usepackage{color}
\usepackage{amssymb}
\usepackage{pifont}
\usepackage{latexsym}
\usepackage{bbm}
\usepackage{subfigure}
\usepackage{float}
\usepackage{epsfig}
\usepackage{psfrag}
\usepackage{natbib, hyperref}
\usepackage{bm}
\usepackage{amsthm}
\usepackage[normalem]{ulem}
\usepackage{eucal}
\usepackage{mathrsfs}
\usepackage{url}
\usepackage{braket}
\usepackage{multirow}
\usepackage{ulem}
\usepackage{calligra}
\usepackage[T1]{fontenc}
\usepackage[utf8]{inputenc}
\usepackage{newunicodechar}
\usepackage{marvosym}
\usepackage[absolute]{textpos}
\usepackage[cal=cm]{mathalfa}
\usepackage{soul}
\usepackage{gensymb}

\usepackage{color}

\usepackage{hyperref}
\hypersetup{
colorlinks=true,final=true,
        linkcolor=blue,
        citecolor=blue,
        filecolor=blue,
        urlcolor=blue,
}

\begin{document}
\setlength{\abovedisplayskip}{3pt}
\setlength{\belowdisplayskip}{3pt}

\title{Anisotropic planar Hall effects in Bi$_2$Se$_3$/EuS interfaces: Deciphering the role of proximity induced spin canting and topological spin texture}

\author{Juhi Singh}
\affiliation{Department of Physics, Indian Institute of Technology Roorkee, Roorkee 247667, India}
\author{Karthik V. Raman}
\email{kvraman@tifrh.res.in}
\affiliation{Tata Institute of Fundamental Research, Hyderabad, Telengana 500046, India}
\author{Narayan Mohanta}
\email{narayan.mohanta@ph.iitr.ac.in}
\affiliation{Department of Physics, Indian Institute of Technology Roorkee, Roorkee 247667, India}

\begin{abstract}
Proximity coupling of ferromagnetic insulator EuS to the topological insulator Bi$_2$Se$_3$ has been proposed to break time-reversal symmetry near the surface of Bi$_2$Se$_3$, introducing an energy gap or a tilt in the surface Dirac cone. As an inverse proximity effect, strong spin-orbit coupling available in the topological surface states can enhance the Curie temperature of ferromagnetism in EuS largely beyond its bulk value, and also generate a magnetic anisotropy. This can result in a canting of the magnetic moment of Eu ions in a plane perpendicular to the interface. Here, we investigate theoretically electronic transport properties arising from the Bi$_2$Se$_3$/EuS interfaces in the planar Hall geometry. Our analysis, based on a realistic model Hamiltonian and a semi-classical formalism for the Boltzmann transport equation, reveals distinct intriguing features of anisotropic planar Hall conductivity, depending on different scenarios for the canting of the Eu moments: fixed Eu moment canting, and freely-orientable Eu moment in response to the external in-plane magnetic field. The anisotropy in the planar Hall conductivity arises from the asymmetric Berry curvature of the gapped topological surface states. We also explore topological Hall effect of the Dirac surface states, coupled to a skyrmion crystal which can emerge in the EuS due to the interplay of ferromagnetic Heisenberg exchange, interfacial Dzyaloshinskii-Moriya interaction, and perpendicular alignment of the Eu moment. Our study provides new impetus for probing complex interplay between magnetic exchange interactions and topological surface states via anisotropic planar Hall effects.
\end{abstract}

\maketitle

\section{Introduction} 
\vspace{-0.3cm}
Three-dimensional topological insulators, when interfaced with a ferromagnet or a superconductor, are known to generate a wide variety of intriguing phenomena such as quantum anomalous Hall effect \cite{Chang_Science2013, Chang_Rev.Mod.Phys2023, Sun_PRL2019, pan_SciAdv2020, moon_NanoLett2019}, topological magnetoelectric effect \cite{Morimoto_PRB2015, dziom_NatCommun2017} and putative topological superconductivity \cite{wang_Science_2012, shen_npjQuantMat_2017}. Particularly, by interfacing Bi$_2$Se$_3$ with EuS, a ferromagnetic insulator with partially filled 4$f$ orbitals and Curie temperature $T_c$$\sim$$17$~K, a strong proximity effect was observed at the interface. The $T_c$ in EuS film was found to enhance beyond its bulk value~\cite{katmis_Nat2016}, attributed to the large enhancement in exchange coupling in EuS near the interface mediated by the Ruderman-Kittel-Kasuya-Yosida type interactions~\cite{Kim_PRL2017, Subramanian_npjQMat2020}. As an inverse proximity effect, the presence of topological surface states of Bi$_2$Se$_3$, having strong spin-orbit coupling, modify spin ordering in EuS. As the thickness of EuS is increased, the induced out-of-plane magnetic anisotropy turns in-plane~\cite{katmis_Nat2016, Kim_PRL2017, Subramanian_npjQMat2020, Assaf_PRB2015, Wei_PRL2013, Changmin_NatCommun2016}. As a result, the Eu moment in a thin film of EuS acquires a canting out of the interface plane. The magnetic exchange field due to the canted Eu moment lifts the degeneracy of the surface Dirac bands, opening up an anisotropic energy gap in the surface Dirac bands.

Planar Hall effect (PHE), arising from a resistivity anisotropy induced by an applied in-plane magnetic field, was found prominently on the surface of Bi$_2$Se$_3$ at energies near the Dirac point~\cite{Taskin_NatCommun2017}. The origin of this resistivity anisotropy in Bi$_2$Se$_3$ is not very well understood. Berry curvature of bulk conduction bands was theoretically proposed to produce a PHE, giving rise to a standard $\cos{\theta_{in}} \sin{\theta_{in}}$ variation of the planar Hall conductivity $\sigma_{xy}$ with the in-plane applied field angle $\theta_{in}$~\cite{Nandy_SciRep2018}. The in-plane magnetic field shifts the Dirac point on the interfacing Bi$_2$Se$_3$ surface relative to the Dirac point on the bottom surface. Proposals for non-topological origin of the PHE include anisotropic back-scattering by magnetic disorders~\cite{Trushin_PRB2009,Chiba_PRB2017} and by the tilted Dirac cone~\cite{Zheng_PRB2020}. In Bi$_2$Te$_3$, a rise in the PHE with the thickness of the sample was found, implying that the bulk states also contribute to the PHE~\cite{bhardwaj_ApplPhysLett_2021}. Subsequent experiments on (Bi, Sb)$_2$Te$_3$/EuS interfaces reported unconventional PHE which requires the consideration of the large proximity effect at the interface~\cite{Rakhmilevich_PRB2018}. Signatures of topological spin textures such as magnetic skyrmions have also been found in PHE measurements~\cite{Dhavala_2024submission}.

\begin{figure*}[t]
\begin{center}
\vspace{-0mm}
\epsfig{file=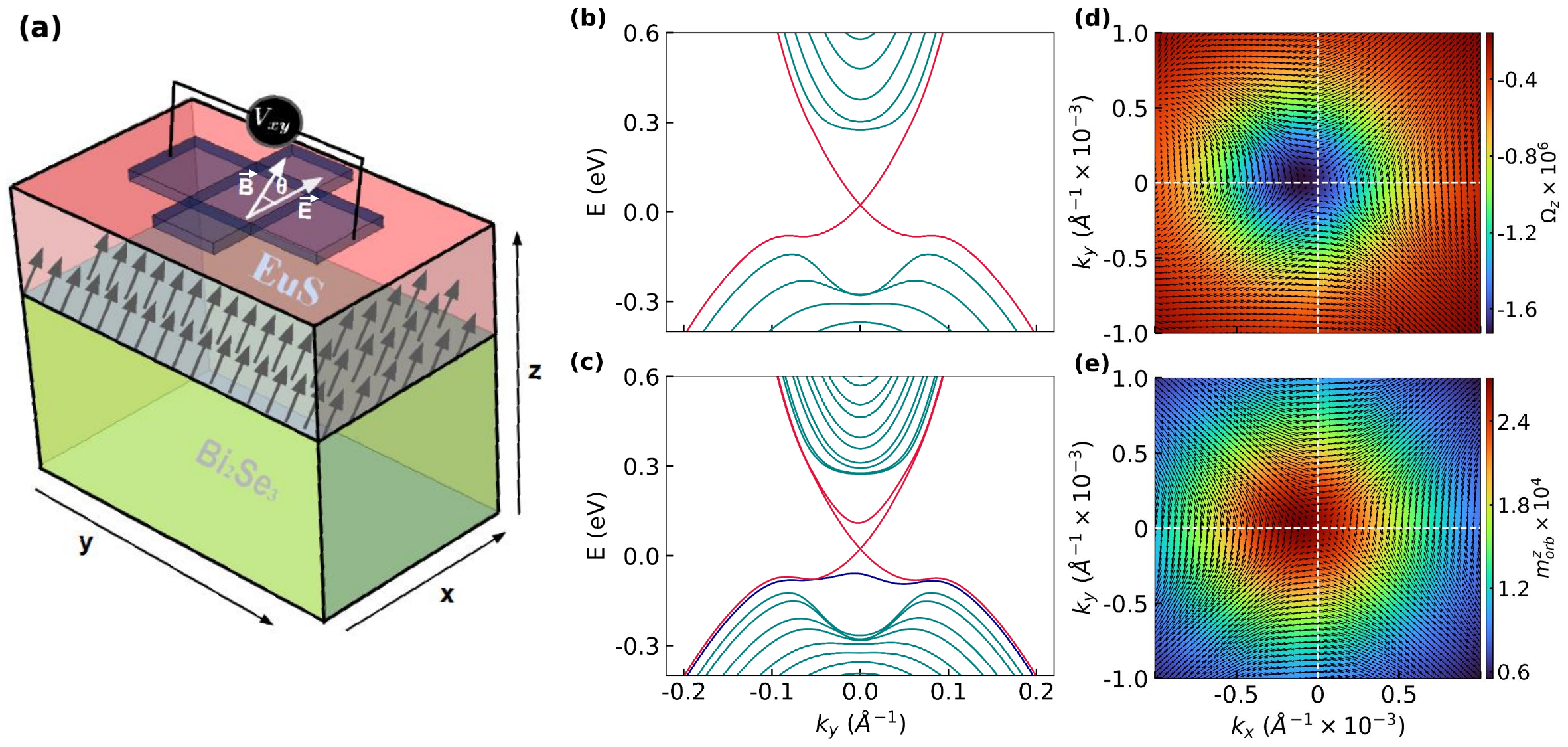,trim=0.0in 0.0in 0.0in 0.0in,clip=false, width=170mm}
\caption{(a) Planar Hall measurement geometry for the Bi$_2$Se$_3$/EuS interface system, considered in this paper. Electric field ${\bf E}$ (due to applied charge current) is oriented along the $x$ axis. The applied magnetic field ${\bf B}$ makes an angle $\theta_{in}$ with ${\bf E}$ in the $x$-$y$ plane. The magnetization in EuS is inclined at an angle $\theta_m$ with the $z$-axis and $\phi_m$ with the $x$ axis in the $x$-$y$ plane. (b) Band dispersion of Bi$_2$Se$_3$ slab of seven quintuple layers ($N\!=\!7$), in the absence of any magnetization field, showing two degenerate Dirac surface bands in red and doubly-degenerate bulk bands in green. (c) Band dispersion of Bi$_2$Se$_3$/EuS interface system with $N\!=\!7$ in Bi$_2$Se$_3$ slab, $m\!=\!0.15 ~{\rm eV}$ (magnetic moment induced at the interface due to proximity effect), $\theta_m = 45\degree$, $B\!=\!2~$T, and $\theta_{in}=45\degree$. (d),(e) Berry curvature and orbital magnetic moment for the lower gapped surface band (blue band in 1(c)) with parameters $m\!=\!6.9~\rm{\mu_B}$, $\theta_m\!=\!45\degree$, $B\!=\!2~$T and $\theta_{in}=45\degree$. In (d) and (e), the colormap shows the $z$-components, while the $x$ and $y$-components are shown by the arrows.}
\label{fig1}
\vspace{-4mm}
\end{center}
\end{figure*}

In this paper, motivated by these recent experimental observations, we investigate the magnetic proximity effect at the Bi$_2$Se$_3$/EuS interface and the influence of the canting of the Eu moment on the resulting PHE. We find anisotropic PHE of different features, depending on the interplay of the Eu moment and the applied in-plane magnetic field. We use a realistic slab Hamiltonian for the interface system to calculate the planar Hall conductivity, within the framework of semi-classical Boltzmann formalism, originating from the Berry curvature of the anisotropically-gapped Dirac cone on the interfacing Bi$_2$Se$_3$ surface. We find that the PHE appears predominantly within a field range between two critical values. Furthermore, we investigate Hall response from magnetic skyrmions which can emerge at the interface when the Eu moment anisotropy is perpendicular to the interface plane and give rise to a finite scalar spin chirality, another source of the PHE. In this case, we find that with increasing the in-plane magnetic field amplitude, there is a gradual transition from the skyrmion crystal phase to the in-plane ferromagnetic phase and the topological Hall conductivity decreases monotonically.

The remainder of this paper is organized as follows.
In Sec. II, we describe our theoretical model for the Bi$_2$Se$_3$/EuS interface and the method to calculate the planar Hall conductivities within semi-classical Boltzmann transport formalism. In Sec. III, we present our numerical results of the anisotropic PHE which can originate from the momentum-space Berry curvature, for different orientation of the Eu moment. In Sec. IV, we show the planar topological Hall signature at the interface due to skyrmion spin texture on the EuS side of the interface. In Sec. V, we provide further discussion on our results regarding the connection to possible experimental observation, and summarize our results.

\section{Model and Method}
\vspace{-0.1cm}
The planar Hall geometry for the Bi$_2$Se$_3$/EuS interface, with canted spin moments in EuS, is shown schematically in Fig.~\ref{fig1}(a). The interface properties can be accessed via the Hall probes using contacts bored through EuS~\cite{Dhavala_2024submission}. Bi$_2$Se$_3$ appears in rhombohedral crystal structure, with space group $\mathrm{D}^5_{3\mathrm{d}}(\mathrm{R}\Bar{3}\mathrm{m})$, in which a five-layer building block consisting of inequivalent Bi and Se atoms, known as the `quintuple' layer repeats along perpendicular to these layers~\cite{Zhang_NatPhys2009}. We theoretically model Bi$_2$Se$_3$ of a finite thickness by considering a slab geometry, in which open boundary conditions are imposed along the stacking direction of the quintuple layers, and periodic boundary conditions along the other two orthogonal directions within each quintuple layer. Therefore, each quintuple layer of Bi$_2$Se$_3$ can be represented by a two-dimensional momentum-space Hamiltonian, and these Hamiltonians are coupled along the stacking direction (considered to be the $z$ axis) in real coordinate space. To incorporate the magnetic proximity effect due to EuS thin film and the influence of the applied in-plane magnetic field, Zeeman exchange coupling is introduced in the Hamiltonian for the topmost layer of ${\rm Bi_2 Se_3}$.

The total slab Hamiltonian for $N$ quintuple layers is given by~\cite{Mohanta_SciRep2017}
\begin{align}
    H({\bf k}_{\parallel}) = \begin{pmatrix}
        H_0 + H_B & H_1 &  &   &   &  &  \\
        H^{\dag}_1 & H_0 & H_1 &  &  &  &  \\
          & H^{\dag}_1 & H_0 & H_1 &   &  &  \\
          & &  .&   &   &  &  \\
          &  &   &.   &  &  &  \\ 
          &  &   &   &.  &  &  \\  
          &  &  &  &   H^{\dag}_1 & H_0 & H_1\\
          &  &  & & &   H^{\dag}_1 & H_0 \\
    \end{pmatrix},
    \label{H_total}
\end{align}
\noindent which is written in the basis $\ket{P1^+_z, \uparrow, l_z}$, $\ket{P2^-_z, \uparrow, l_z}$, $\ket{P1^+_z, \downarrow, l_z}$, $\ket{P2^-_z, \downarrow, l_z}$, where $P1^{\pm}_z$ and $P2^{\pm}_z$ denote orbitals originated from the $p$ orbitals of Bi and Se~\cite{Liu_PRB2010}, ${\bf k}_{\parallel}\! \equiv \!(k_x,k_y)$, $(\downarrow,\uparrow)$ denotes the electron spin quantization states, and $l_z$ is the layer index which varies between 0 and $N$. The Hamiltonians $H_0$, $H_1$ and $H_B$ describe, respectively, a single quintuple layer, coupling between two neighboring quintuple layers, and Zeeman exchange coupling on the top surface layer due to proximity effect of ferromagnetic EuS and due to the applied in-plane magnetic field. These Hamiltonians can be written as

\begin{align}
 \hspace{-0.7cm} &H_0  =\begin{pmatrix}
  \epsilon_0+M & 0 & 0 & A_0{\bf k_-}  \\ 
        0 & \epsilon_0-M & A_0{\bf k_-} & 0 \\ 
        0 & A_0{\bf k_+} & \epsilon_0+M & 0\\
        A_0{\bf k_+} & 0 & 0 & \epsilon_0-M \\
\end{pmatrix}, 
\end{align}

\begin{align}
&H_1  =\begin{pmatrix}
  -M_1 -C_1 & i B_0/2 & 0 & 0 \\ 
        i B_0/2 & M_1 - C_1 & 0 & 0 \\ 
        0 & 0 & -M_1 - C_1 & -i B_0/2\\
        0 & 0 & -i B_0/2 & M_1 - C_1 \\
\end{pmatrix}, 
\end{align}

\begin{align}
&H_B = \begin{pmatrix}
        h_z & 0 & h_x - ih_y & 0 \\
        0 & h_z & 0 & h_x - ih_y \\
        h_x + ih_y & 0 & -h_z & 0\\
        0 & h_x + ih_y & 0 & -h_z \\
     \end{pmatrix},  
\end{align}
\\where $\epsilon_0(\mathbf{k}_{\parallel}) \!=\! C_0 + 2C_1 + 2C_2 (2-\cos(k_xa)-\cos(k_ya))$, $M_0(\mathbf{k}_{\parallel}) \!=\! M_0 + 2M_1 + 2M_2 (2-\cos(k_xa)-\cos(k_ya))$, $a$ is the lattice constant, ${\bf k_{\pm}}=\sin{(k_xa)}\pm i\sin{(k_ya)}$, $h_x \!=\! m_x + g\mu_B B_x/2$,  $h_y \!=\! m_y + g\mu_B B_y/2$, and $h_z \!=\! m_z$. $ m_x \!=\! m \sin\theta_m \cos\phi_m$, $ m_y \!=\! m \sin\theta_m \sin\phi_m$, $m_z \!=\! m \cos\theta_m$,  $B_x \!=\! B \cos\theta_{in}$, and $B_y \!=\! B \sin\theta_{in}$, ${\bf m} \! \equiv (m_x,m_y,m_z)$ represents the $\mathrm{EuS}$ magnetic moment which is assumed to be canted at a polar angle $\theta_m$ with the $z$-axis and azimuithal angle $\phi_m$ with the $x$-axis on the $x$-$y$ (interface) plane, and $B$ is the magnitude of the external in-plane magnetic field applied at an angle $\theta_{in}$ with the $x$-axis on the $x$-$y$ plane. 
The parameter values used in this paper are $A_0 = 0.8$~eV, $B_0 = 0.32$~eV, $C_0 = -0.0083$~eV, $C_1 = 0.024$~eV, $C_2 = 1.77$~eV, $M_0 = -0.28$~eV, $M_1 = 0.216$~eV, $M_2 = 2.6$~eV, $g=20$, and $a = 4.14 {\AA}$~\cite{Mohanta_SciRep2017}. Fig.~\ref{fig1}(b) and (c) show the band dispersions of a Bi$_2$Se$_3$ slab of seven quintuple layers, without and with the Zeeman exchange coupling due to the Eu moment and external applied magnetic field. The combined effect of the applied in-plane magnetic field and the canted Eu moment, anisotropically lifts the surface bands on the top surface of Bi$_2$Se$_3$, as shown in Fig.~\ref{fig1}(c).


\begin{figure*}[t]
\begin{center}
\vspace{-0mm}
\epsfig{file=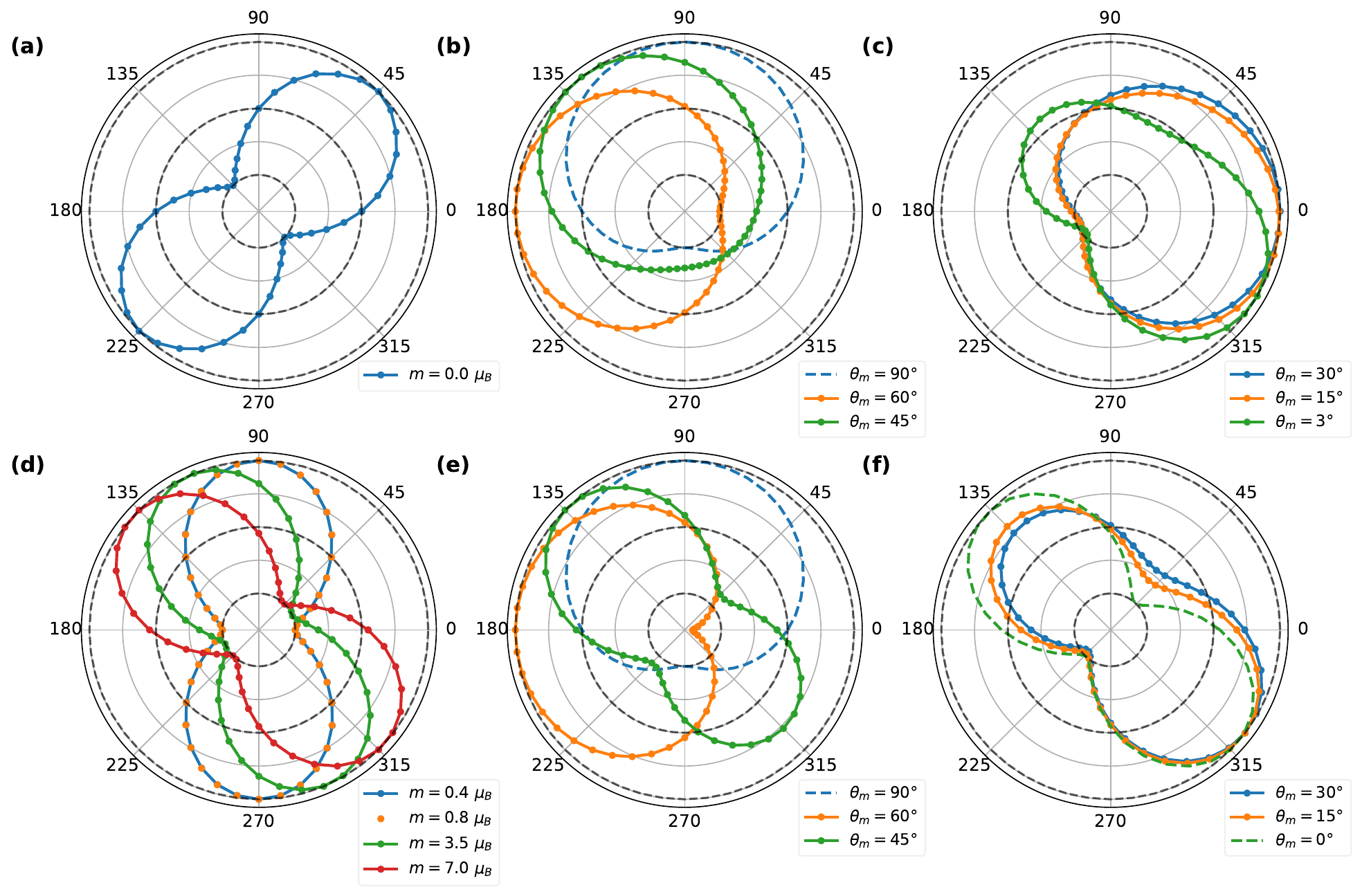,trim=0.0in 0.0in 0.0in 0.0in,clip=false, width=160mm}
\caption{Planar Hall conductivity, $\sigma_{xy}$ normalized with respect to its maximum value, is plotted for (a) $m=0.0~\mu_B$, (b)-(c) $m=6.9~\mu_B$, $\phi_m=0\degree$, (d) $m=6.9$ to $0.4\mu_B$, $\theta_m = \phi_m=0\degree$, and (e)-(f) $m=6.9~\mu_B$, $\phi_m=0\degree$. The external in-plane field $B$ is fixed to $30~{\rm mT}$ for plots (a)-(d) and to $B=0.5~{\rm T}$ for (e)-(f). The polar canting angle $\theta_m$ is varied from $90\degree$ (in-plane) to $0\degree$ (out-of-plane). The PHC is symmetric {\it i.e.}, $\sigma_{xy}(B) = \sigma_{xy}(-B)$ with $\pi$ period and follows a $\sin \theta_{in} \cos \theta_{in}$ for $m=0.0~\mu_B$, while an angular shift is observed for $(m_x, my, m_z) \equiv (0,0,m)$ as shown in (a) and (d). For cases with Eu moment aligned in either $x$ or $y$ direction, antisymmetric behavior $(\sigma_{xy}(B) = -\sigma_{xy}(-B))$ is observed with $2\pi$ period. For $0\degree <\theta_m <90\degree$, the PHC reveals an anisotropic behavior with $A \sin\theta_{in} \cos\theta_{in} + B\cos\theta_{in}$ dependence, where $A,B \in \mathbb{R}$. }
\label{fig2}
\vspace{-0mm}
\end{center}
\end{figure*}

\begin{figure*}[t]
\begin{center}
\vspace{-0mm}
\epsfig{file=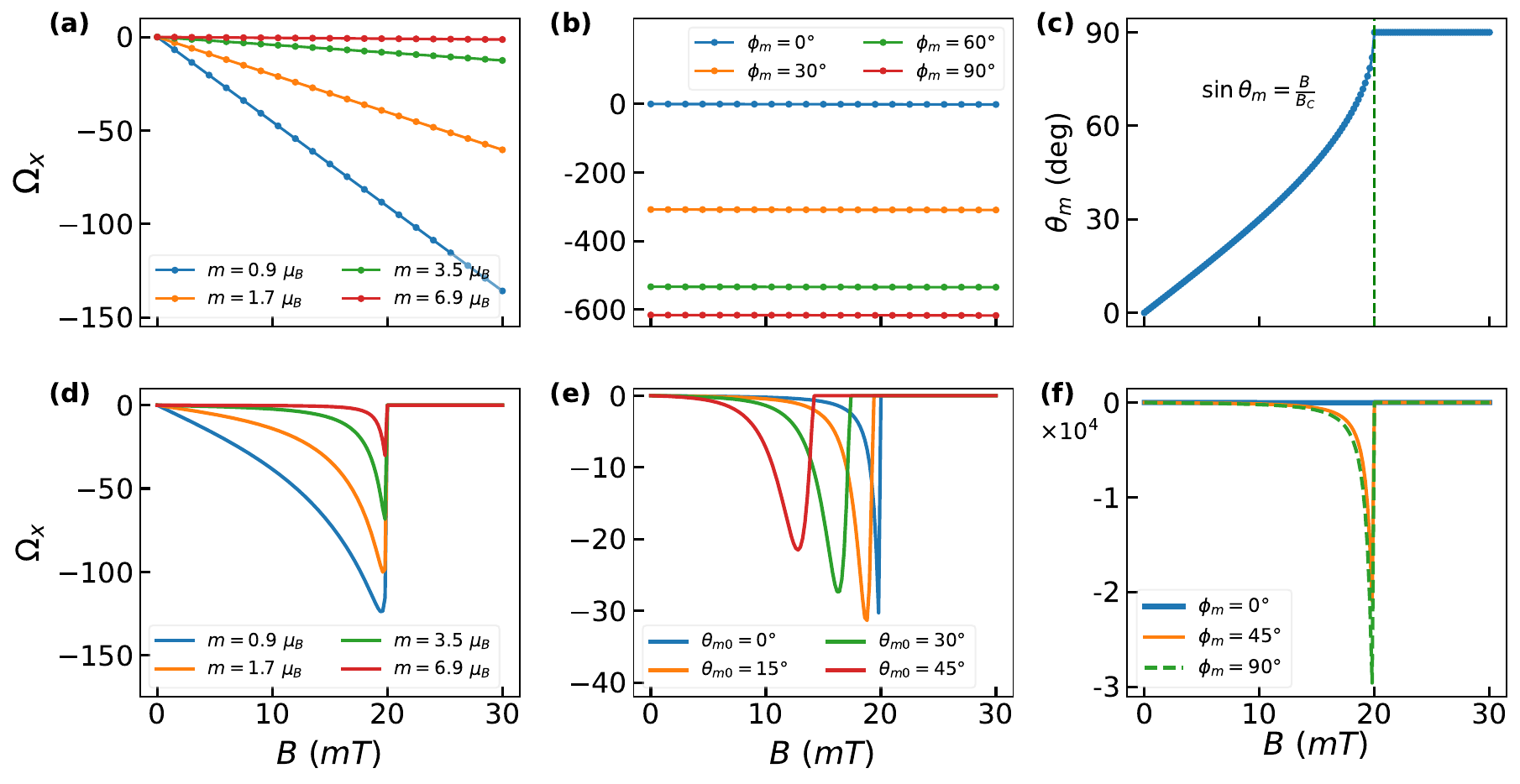,trim=0.0in 0.0in 0.0in 0.0in,clip=false, width=180mm}
\caption{(a)-(b) The x-component of Berry curvature $(\Omega_x)$ at $(k_x,k_y)=(0,0)$ plotted as a function of in-plane field $B$ for different values of Eu moment $m$ and azimuthal canting angle $\phi_m$. For (a) $\phi_m=0\degree$ and the external field angle $\theta_{in}$ and the polar canting angle $\theta_{m}$ for (a) and (b) are both fixed at $45\degree$. (c) $\theta_m$ plotted as a function of the in-plane magnetic field amplitude $B$, $\theta_m=\sin^{-1}(B/B_c)$ as used in the calculations. For $B\geq B_c (=20~{\rm mT})$, $\theta_m ~{\rm equals}~ 90\degree$ and the Eu moment turns in-plane. (d)-(f) depicts the variation of the $x$- component of Berry curvature $(\Omega_x)$ at $(k_x,k_y)=(0,0)$ with $B_{in}$ for different values of the Eu moment $m$, initial canting angle $\theta_{m0}$ and azimuthal canting angle $\phi_m$. The Berry curvature is maximum at a field value close to $B_{c}$ and then drops to zero at $B_{in}\!=\!B_{c}$. In these plots, the polar canting angle of the Eu moment varies with the applied field {\textit i.e.} $\sin \theta_m \!=\! B/B_c$.}
\label{omega}
\vspace{-0mm}
\end{center}
\end{figure*}


\begin{figure*}[t]
\begin{center}
\vspace{-0mm}
\epsfig{file=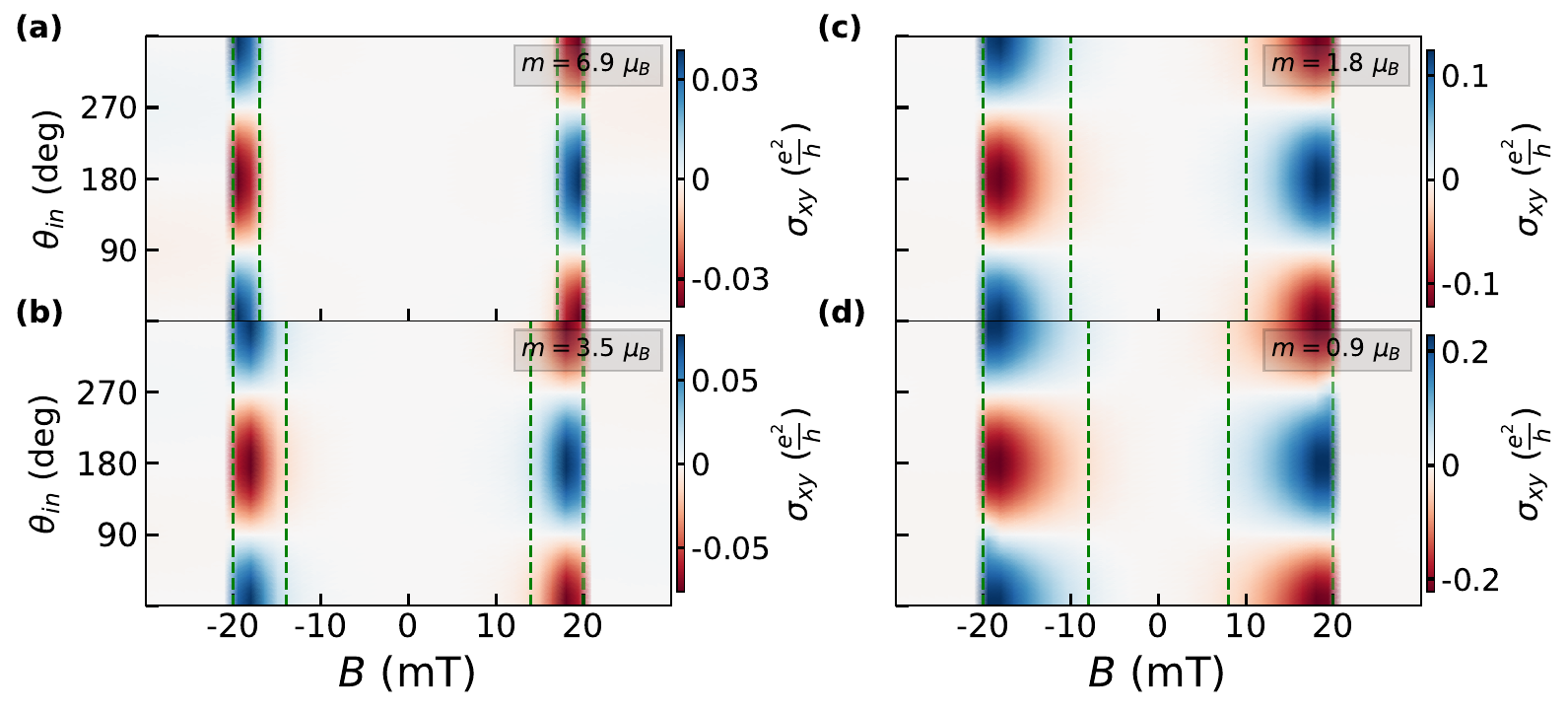,trim=0.0in 0.0in 0.0in 0.0in,clip=false, width=180mm}
\caption{Planar Hall conductivity, $\sigma_{xy}$ plotted for different magnitudes of Eu moment, (a) $m=6.9~\mu_B$, (b) $m=3.5~\mu_B$, (c) $m=1.8~\mu_B$, and (d) $m=0.9~\mu_B$. The polar canting angle $\theta_m$ follows $\theta_m = \sin^{-1}(B/B_c)$ relation and equals $90\degree$ for $B \geq B_c$. The azimuthal canting angle $\phi_m$ is fixed to $0\degree$ and the initial configuration is assumed to be in the perpendicular direction to the interface i.e., $\theta_{m0}=0\degree$. The second critical field $B_c$ is set to $20 ~{\rm mT}$. The first critical field, $B_{c^{\prime}}$ decreases with decreasing $m$ and the width $\Delta B = |B_c-B_{c^{\prime}}|$ widens. Other parameters are $N=7$, $T=24{\rm K}$, and $\tau = 10^{-13}s$.}
\label{fig4}
\vspace{-0mm}
\end{center}
\end{figure*}

The planar Hall conductivity (PHC) is obtained within the semi-classical Boltzmann formalism (see Appendix C for details) as
\begin{equation}\label{s_yx}
\begin{split}
\sigma_{xy} =& - \frac{e^2}{4 \pi^2} \int d{\bf k} ~ D \tau ~\Big[\tilde{v}_y \tilde{v}_x + \frac{e B \cos \theta_{in} ~ \tilde{v}_y }{\hbar} (\bf{\tilde {v}_k} \cdot \Omega(\bf{k})) \\ 
&+\frac{eB \sin \theta_{in} ~ \tilde{v}_x }{\hbar} ({\bf \tilde{v}_k}  \cdot {\bf \Omega(k)}) \\
&+\frac{e^2 B^2 \cos \theta_{in} \sin \theta_{in}}{\hbar^2} ({\bf \tilde{v}_k} \cdot {\bf \Omega( k)})^2\Big] \partial_{\tilde{\epsilon}} \tilde{f}_{eq},
\end{split}
\end{equation}
\noindent where ${\bf \tilde {v}_k}\equiv(\tilde{v}_x,\tilde{v}_y,0)$ is the velocity of electrons at the surface of Bi$_2$Se$_3$ which interfaces EuS, modified by the orbital magnetic moment (OMM), ${\bf \Omega(k)}\equiv(\Omega_x, \Omega_y, \Omega_z)$ is the Berry curvature, and $\tilde{f}_{eq}$ represents equilibrium Fermi-Dirac distribution with modified energy(see Appendices C-D for details). Here, the transverse Hall conductivity $\sigma_{xy}$ does not satisfy Onsager's reciprocity relation i.e., $\sigma_{xy} = -\sigma_{yx}$, hence it can be utilized to characterize the planar Hall effect~\cite{Battilomo_PRR2021,Wang_PRL2024,Xiang_PRB2024}.
The first term on the right-hand side of Eq.~(\ref{s_yx}), i.e., $\tilde{v}_x \tilde{v}_y$, is the usual planar Hall contribution due to the presence of in-plane electric and magnetic fields. Due to the magnetic proximity effect of EuS film on the upper surface of Bi$_2$Se$_3$, a finite Berry curvature is induced in the surface band, allowing the other three terms to contribute considerably.  When Zeeman coupling is perpendicular to the interface i.e., in the $z$-direction, the z-component of Berry curvature, $\Omega_z$ is circularly symmetric centered at $(k_x,k_y)=(0,0)$ in the $k$-space. With the introduction of in-plane field components, the gapped Dirac point on the interfacing Bi$_2$Se$_3$ surface is shifted in momentum compared to the gap-less Dirac point on the opposite Bi$_2$Se$_3$ surface. Consequently, the maximum of $\Omega_z$ shifts from the center of the Brillouin zone. Fig.~\ref{fig1}(d) and (e) show the Berry curvature and OMM of the anisotropically-gapped surface band, shown in Fig.~\ref{fig1}(c). We note that the Berry curvature and the OMM show textures analogous to anti-meron.

\section{Berry curvature driven planar Hall effect}
The planar Hall effect observed in a topological insulator is symmetric with respect to the applied in-plane magnetic field and follows a $\sin \theta_{in} \cos \theta_{in}$ dependence as shown in Fig.~\ref{fig2}(a). In this case, the only contribution is from the first term in the Eq.~(\ref{s_yx}) i.e., the product of the modified velocities. However, at the $\rm{Bi}_2 \rm{Se}_3/\rm{EuS}$ interfaces, planar Hall conductivity will depend on the orientation of the Eu moment at the interface, which is governed by the magnitude and direction of the external in-plane magnetic field. In the following discussion, we have considered different configurations of the Eu moment to explore different possible planar Hall signatures at the interface.

\subsection{Fixed orientation of the Eu moment}
In this case, the Eu moment is assumed to be independent of the external in-plane magnetic field. We consider that the azimuthal canting angle $\phi_m$ of the Eu moment is fixed at $0\degree$ \textit{i.e.} the Eu moment is on the $x-z$ plane, while the polar canting angle is decreased from $90\degree$ (in-plane) gradually with increasing the in-plane field to $0\degree$ (out-of-plane). Also, the magnitude of the Eu moment is fixed at $6.9~\mu_B$, the magnetic moment of bulk EuS as reported in Ref.~\cite{Kim_PRL2017}. Fig.~\ref{fig2}(b)-(c), depicts the PHC for $\theta_m$ varying from $90\degree$ (Eu moment aligned in the $x$-direction) to $3\degree$, an anti-symmetric contribution is observed with $2\pi$ periodicity, which transitions to symmetric contribution with $\pi$ periodicity for $\theta_m = 0\degree$ as shown in Fig.~\ref{fig2}(d). If the Eu moment is considered on the $y-z$ plane i.e.,  $\phi_m=90\degree$, the dependence shifts by $90\degree$ from $\sin\theta_{in}$ to $\sin (90 + \theta_{in}) = \cos \theta_{in}$. Since $x$ and $y$ directions are giving similar results with a $90\degree$ phase shift. Therefore, we have considered the easy-axis alignment in the $x$-$z$ plane $(\phi_m=0)$ for further results. From the numerical results, we find that the $\theta_{in}$ dependence of the PHC changes from $\sin\theta_{in}$ for $\theta_m=90\degree$ to $-\cos\theta_{in}$ for $\theta_m=60\degree$, and $\cos\theta_{in}$ at $\theta_m=30\degree$. Furthermore, for $\theta_{m}\leq 5\degree$, the PHC behaves as $ \cos \theta_{in} - \sin \theta_{in} \cos \theta_{in}$ and eventually follows $- \sin \theta_{in} \cos \theta_{in}$ dependence for $\theta_m = 0\degree$. On generalizing the results for $0\degree<\theta_m \leq 90\degree$, the PHC reveals an anisotropic behavior with $A \sin\theta_{in} \cos\theta_{in} + B\cos\theta_{in}$ dependence, where $A,B \in \mathbb{R}$. 
The transition from $\sin \theta_{in}$ to $-\sin \theta_{in} \cos \theta_{in}$ is visible at very small $\theta_m$ for low in-plane magnetic field. However, if the in-plane magnetic field is increased to $0.5~\mathrm{T}$, this transition begins at higher polar canting angles, as shown in Fig.~\ref{fig2}(e) and (f). This happens because, at a higher in-plane field, the effect of the magnetic moment due to EuS is suppressed by the external field. Now, varying the magnitude of Eu moment $m$ from $0$ to $6.9~\mu_B$ for $\theta_m=\phi_m=0\degree$, the PHC shifts from $\sin \theta_{in} \cos \theta_{in}$ to $\sin (\theta_{in}+\theta') \cos (\theta_{in}+\theta')$, where $\theta'$ represents the angular shift. This shift in angle $\theta'$ increases with $m$ and equals $90\degree$ for $m=6.9~\mu_B$ as shown in Fig.~\ref{fig2}(d). The computed PHC discussed above is normalized with respect to its maximum value. The anisotropy in the PHC is a direct consequence of the anomalous velocity introduced by the anisotropic in-plane components due to Eu moment and the external magnetic field. 

To understand the variation in PHC with changing different parameters, we may refer to the change in the magnitude of different components of the Berry curvature ${\bf \Omega(k)}$. Fig.~\ref{omega}(a) shows the variation of the $x$-component of the Berry curvature $(\Omega_x)$ with the magnitude of Eu moment $m$. We find that $\Omega_x$ increases linearly with the in-plane field and the slope of this linear plot depends inversely on $m$. Therefore, the PHC also varies inversely with $m$. Also, $\Omega_x$ increases with increasing the azimuthal canting angle $\phi_m$ as shown in Fig.~\ref{omega}(b). However, $\Omega_y$ shows an inverse relation, leading to no significant change in the magnitude of the PHC on varying $\phi_m$. 

The above results indicate that the anisotropy in the PHC with respect to the applied in-plane field is due to the orientation of the Eu moment, while the magnitude $m$ of the Eu moment is responsible for angular shift $\theta^{\prime}$.

\begin{figure}[t]
\begin{center}
\vspace{-0mm}
\epsfig{file=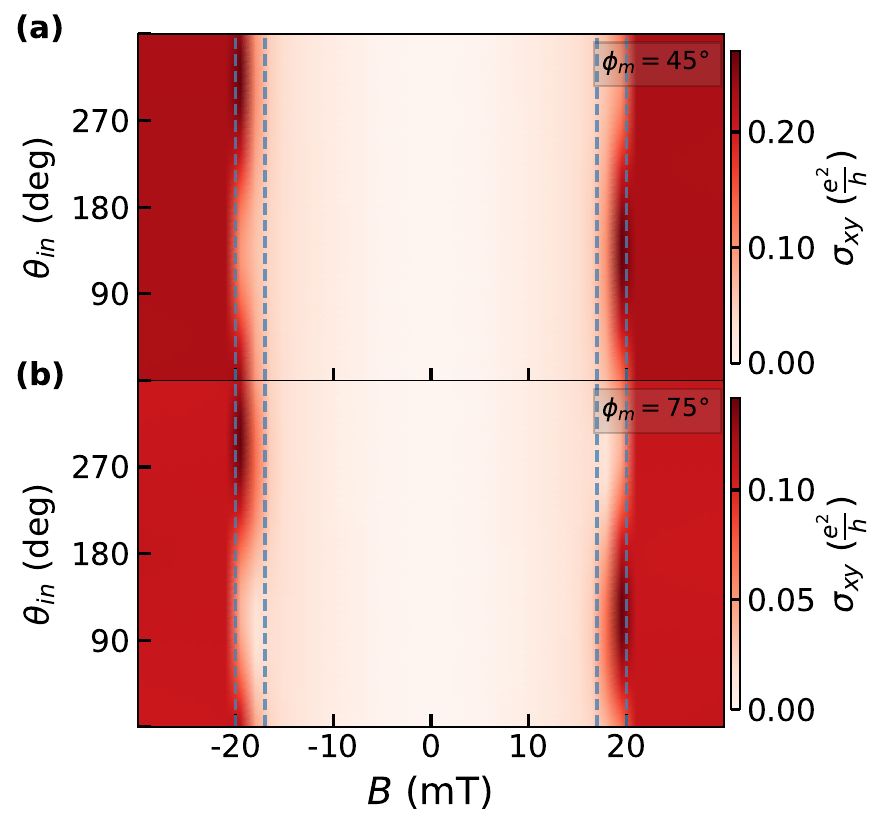,trim=0.0in 0.0in 0.0in 0.0in,clip=false, width=86mm}
\caption{Planar Hall conductivity $\sigma_{xy}$, plotted in the plane of the magnitude $B$ and the angle $\theta_{in}$ of the external in-plane magnetic field, for the cases (a) $\phi_m \!=\! 45\degree$, and (b) $\phi_m\!=\!75\degree$. The initial canting angle of the Eu moment and the critical magnetic field, considered in both cases, are $\theta_{m0}\!=\!0\degree$, and $B_c\!=\!20~\mathrm{mT}$, respectively. The magnetic moment is fixed to $m=6.9~\mu_B$ and the polar canting angle follows the $\sin \theta_m = B/B_c$ relation.}
\label{fig5}
\vspace{-1mm}
\end{center}
\end{figure}

\subsection{Eu moment reorienting with in-plane field}
Considering the general configuration for the magnetic moment of Eu ions ${\bf m} \! \equiv (m_x,m_y,m_z)$ canted at a polar angle $\theta_m$ with the $z$-axis and azimuthal angle $\phi_m$ with the $x$-axis on the $x$-$y$ plane, the energy-minimized configuration is given by (see Eq.~(\ref{etot}) in Appendix B)  $\phi_m = \theta_{in}$ and
\begin{equation} \label{thetam}
\begin{split}
\sin \theta_m & =\frac{StM_sB}{2Au^2m} = \frac{B}{B_c}, ~~~  ~B<  B_c  \\
 & = 1,  \hspace{2.5cm}  B \geq B_c
\end{split}
\end{equation}

where $t$ is the thickness of $\rm{EuS}$ film, $M_s$ is the saturation magnetization, $S$ is the area of the two-dimensional system of topological surface states, $u$ is the exchange coupling strength, and $A=S(\sqrt{N_c}-\sqrt{|2N_e-N_c|})/(2\sqrt{\pi}\hbar |v_f|)$ with $N_c$ and $N_e$ being the cutoff and total electron densities respectively. The above relation indicates that the Eu moment turns in-plane when the field amplitude $B$ exceeds the critical value $B_c$, which is considered as a parameter in our discussion below.

Fig.~\ref{fig4}(a) shows the variation in PHC with the above setting of the polar canting angle $\theta_m$ and $\phi_m$ is fixed to $0\degree$. We observe following three regimes of the in-plane field: (\romannumeral1\relax) $0 \leq |B|<|B_{c^{\prime}}|$, (\romannumeral2\relax) $|B_{c^{\prime}}| \leq |B| \leq |B_{c}|$, and (\romannumeral3\relax) $|B_c|<|B|$, where $B_{c^{\prime}}$ and $B_{c}$ are first and second critical fields, respectively. The first critical field, $B_{c^{\prime}}$ is the field value at which the Berry curvature becomes significant to contribute to the Hall conductivity. The second critical field $B_c$, represents the threshold field strength beyond which the Eu moment transitions to an in-plane orientation. 

The emergence of critical fields can be understood by looking at the behavior of the components of the Berry curvature ${\bf \Omega(k)}$. For the lower gapped surface band, the x-component of berry curvature $\Omega_x$ at $(k_x, k_y) \equiv (0,0)$ is plotted as a function of the in-plane field in Fig.~\ref{omega}(d)-(f).  It is observed that $\Omega_x$ exhibits an increasing pattern with $B$, attains a peak value, and subsequently diminishes to zero for $B \geq B_c$. In the following discussion the second critical field $B_c$ is fixed to $20~$mT, while other parameters are varied to observe the changes in $B_{c^{\prime}}$, $\Delta B = |B_c - B_{c^{\prime}}|$, and the PHC. 

\begin{figure}[t]
\begin{center}
\vspace{-0mm}
\epsfig{file=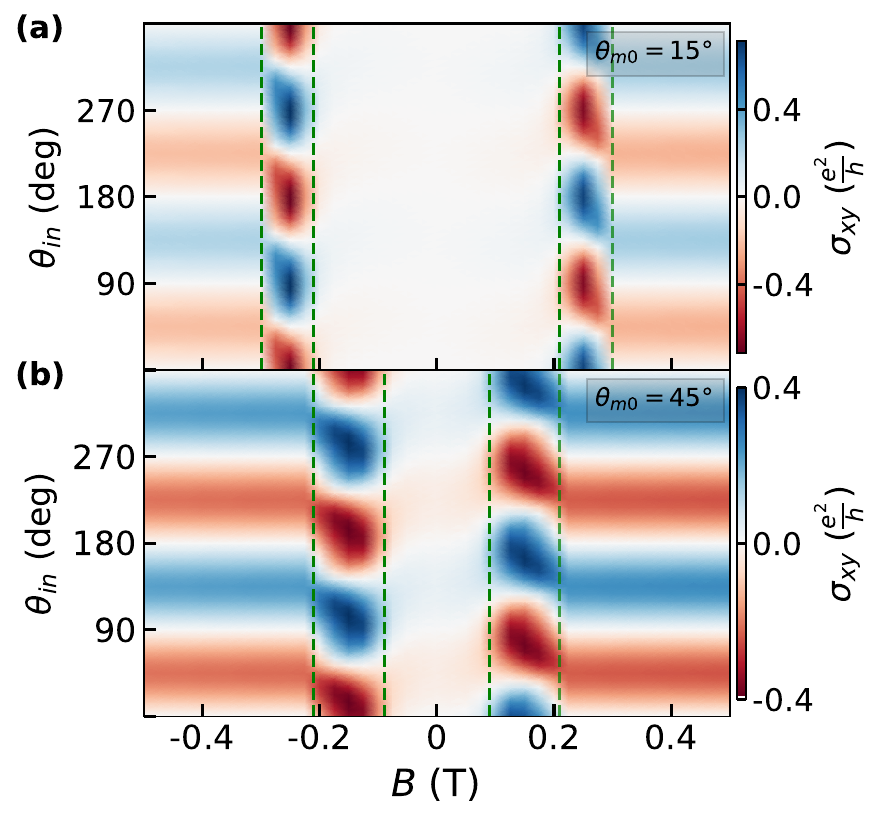,trim=0.0in 0.0in 0.0in 0.0in,clip=false, width=86mm}
\caption{Planar Hall conductivity $\sigma_{xy}$, plotted in the plane of the magnitude $B$ and the angle $\theta_{in}$ of the external in-plane magnetic field, for the cases (a) $\theta_{m0} \!=\! 15\degree$, and (b) $\theta_{m0}\!=\!45\degree$. The azimuthal canting angle and the critical magnetic field, considered in both cases, are $\phi_{m}\!=\!\theta_{in}$, and $B_c\!=\!0.3~\mathrm{T}$, respectively. The magnetic moment is fixed to $m=6.9~\mu_B$ and the polar canting angle follows the $\sin \theta_m = B/B_c$ relation.}
\label{fig6}
\vspace{-1mm}
\end{center}
\end{figure}

\vspace{0.2cm}
\noindent \textit{Changing magnitude of Eu moment}

The magnitude of the Eu moment is directly proportional to the thickness of the interfaced thin film EuS, making it an important parameter for our study. For lower values of the canted Eu moment, with a finite component along the $z$-direction, the PHC gives a larger value as shown in Fig.~\ref{fig4}. The increase in the PHC with a decrease in magnetic moment is a direct consequence of a smaller gap opening at lower field values, which leads to large Berry curvature. Fig.~\ref{omega}(d) shows the variation of the $x$-component of the Berry curvature with the in-plane magnetic field. The maximum value attained by $\Omega_x$ increases with a decrease in the magnitude of the Eu moment. The enhanced first critical field value can be understood from the increased field range within which $\Omega_x$ remains large. Therefore, we can express the dependence as 
$$|B_{c^{\prime}}| \propto |B_{c}|\propto {m} ~~~{\rm and} ~~~|\Delta B|=|B_c-B_{c^{\prime}}| \propto \frac{1}{m}$$

\vspace{0.2cm}

\noindent \textit{Changing azimuthal canting angle}

As shown in Fig.~\ref{omega}(f) varying the azimuthal canting angle does not have a significant effect on the critical fields but can affect the behavior of the PHC significantly. To discuss this aspect, we consider two cases, as mentioned below.\\

\noindent \textit{Case (a): $\phi_m =$ constant}

As shown in Fig.~\ref{fig4}, the PHC for  $\phi_m=0\degree$ is dominant in the second regime i.e., for $|B_{c^{\prime}}| \leq |B| \leq |B_{c}|$ and follows a $\pm \cos \theta_{in}$ dependence for $\mp B$. Also, it has $2\pi$ periodicity with antisymmetric behavior with respect to the external in-plane field.
The behavior of the PHC in this case is similar to the case where the Eu moment does not change with the application of the in-plane field. This can be understood by the fact that, when $B \to 0$, the polar canting angle is nearly $0\degree$ and with increasing value of $|B|$, $\theta_m$ also increases. In the second regime, $\theta_{m}$ nearly equals $60\degree$. Therefore, we get a $-\cos\theta_{in}$ dependence.
 However, if $\phi_m$ is assumed to be $90\degree$, $-\sin \theta_{in}$ dependence will be observed. The $\cos\theta_{in} \tilde{v}_y$ and $\sin\theta_{in} \tilde{v}_x$ terms on the RHS of Eq.~(\ref{s_yx}) are responsible for the above behavior and it remains consistent even at higher magnetic fields.

When $0\degree < \phi_m < 90\degree$, both $\sin \theta_{in} \tilde{v}_x$ and $\cos \theta_{in} \tilde{v}_y$ terms contribute and the PHC follows a linear combination of $\sin\theta_{in}$ and $\cos \theta_{in}$. Fig.~\ref{fig5} depicts the behavior of the PHC for $\phi_m=45\degree$ and $75\degree$. We find that the PHC increases from zero to a finite value as the in-plane field magnitude increases to the first critical field. However, it remains constant with the in-plane field angle for $|B|<|B_{c^{\prime}}|$. In the $|B_{c^{\prime}}| \leq |B| \leq |B_c|$ range, anisotropic PHC is observed with $C\cos\theta_{in} + D \sin\theta_{in}$ dependence, where $C, D \in \mathbb{R}$. 

Also, for $|B| > |B_c|$ the PHC is constant and non-zero, unlike in the previous case; this is because $m_y \neq 0$ and the $x$ and the $y$ components of the Eu moment add to the Zeeman exchange due to the external in-plane field.    

\noindent \textit{Case (b): $\phi_m = \theta_{in}$}

The azimuthal canting angle $\phi_m$ is now assumed to vary along with the in-plane field angle $\theta_{in}$. The second critical field $B_c$ is taken to be $0.3~$T for this case to understand the behavior at a higher magnetic field. As shown in Fig.~\ref{fig6}(a), the PHC is antisymmetric in the regime $|B_{c^{\prime}}| \leq |B| \leq |B_{c}|$, with $\pi$ period, and follows a $\sin(\theta_{in}+\delta) \cos(\theta_{in}+\delta)$ dependence, where $\delta$ represents the angular shift in the PHC. For smaller critical field values $\delta \to 0\degree$, the antisymmetric behavior transitions to symmetric, without any change in periodicity. However, to preserve the antisymmetric behavior at smaller field values, the magnitude of Eu moment $m$ should be lowered. For the regime $|B| > |B_c|$ in Fig.~\ref{fig6}(a), $\sin \theta_{in} \cos \theta_{in}$ dependence is observed, not affected by further increase in the field value. This happens because $m_z$ becomes zero, and $m_x$ and $m_y$ become constant for a field larger than the second critical field. Since the magnitude of the Eu moment is much larger than the in-plane field, we do not see any effect of increasing the in-plane field on the PHC.

\begin{table*}
    \caption{Summary of the results discussed in Sec III. (A) Magnetic moment due to EuS is fixed and does not change its orientation with the external in-plane field. For different combinations of polar and azimuthal canting angles the change in periodicity and symmetry of the PHC is mentioned. (B) Eu moment reorients itself with the external field. The PHC is dominant in the range $|B_{c^{\prime}}|<|B|<|B_{c}|$ of external magnetic field. The variation in the nature of the PHC with changing the azimuthal canting angle $\phi_m$, the initial canting angle $\theta_{m0}$, and the magnitude of the Eu moment $m$ is summarized.}
    \vspace{0.2cm}
    \begin{tabular}{|m {3.5em}| m {6.5em}|m {4em}|m {11.5em}| m {5em}|m{16em}|} \hline 
    \multicolumn{6}{|c|}{A. Fixed Orientation of Eu moment} \\ \hline
    \multicolumn{3}{|c|}{Eu moment alignment} & \multicolumn{3}{|c|}{Hall conductivity} \\ \hline 
    $m (\mu_B)$ & $\theta_m (\degree)$ & $\phi_m (\degree)$ & $\sigma_{xy}=f(\theta_{in})$ & Period & Symmetry with respect to negative field value\\ \hline
    $0.0$ & - & - & $\sin \theta_{in} \cos \theta_{in}$ & $\pi$ & Symmetric \\ \hline
    $<6.9$ & $0\degree$ & - & $\sin(\theta_{in}+\theta')\cos(\theta_{in}+\theta')$ & $\pi$ & Symmetric \\ \hline
    \multirow{4}{2em}{$6.9$} & $0\degree$ & - & $-\sin \theta_{in} \cos \theta_{in}~ (\theta'=90\degree)$ & $\pi$ & Symmetric\\ \cline{2-6}
     & $90\degree$ & $90\degree$ & $\cos \theta_{in}$ & $2\pi$ & Antisymmetric\\ \cline{2-6}
     & $90 \degree$ & $0 \degree$ & $\sin \theta_{in}$ & $2\pi$ & Antisymmetric\\ \cline{2-6}
     & $0 \degree < \theta_m< 90\degree$ & $0\degree$ & \makecell{$(A \sin\theta_{in} + B) \cos\theta_{in}$, \\where $A,B \in \mathbb{R}$} & $2\pi$ & Anisotropic \\     \hline
    \end{tabular}
     \vspace{0.5cm}
     
    \begin{tabular}{|m {6em}|m {7em}|m {13em}|m {4em}|m {17em}|} \hline 
    \multicolumn{5}{|c|}{\makecell{B. Polar canting angle of Eu moment is free to reorient according to $\theta_{m} = \sin^{-1}(B/B_c) + \theta_{m0}$}} \\ \hline \hline
    \multirow{6}{6em}{\makecell{Varying \\azimuthal \\canting angle \\$\phi_m$ \\$m = 6.9~\mu_B$ \\and $\theta_{m0}=0 \degree$}} & \multirow{2}{4em}{$\phi_m (\degree)$} & \multicolumn{3}{|c|}{\makecell{Hall conductivity in range \\$|B_{c^{\prime}}| \leq |B| \leq |B_{c}|$}}  \\ \cline{3-5}
    & & $\sigma_{xy}=f(\theta_{in})$ & Period & Symmetry with respect to negative field value\\ \cline{2-5}
    & $0 \degree$   & $-\cos\theta_{in}$ & $2\pi$ & Antisymmetric \\ \cline{2-5}
    & $90 \degree$ & $-\sin\theta_{in}$ & $2\pi$ & Antisymmetric \\ \cline{2-5}
    & $0 \degree <\phi_m < 90\degree$ & \makecell{$C\cos\theta_{in}+D\sin\theta_{in}$ \\where $C,D \in \mathbb{R}$} & $2\pi$ & Anisotropic\\ \cline{2-5}
    & $\theta_{in}$ & \makecell{$-\sin(\theta_{in}+\delta) \cos(\theta_{in}+\delta)$} & $\pi$ & Antisymmetric (if $m<6.9~\mu_B$ or $B\sim 0.5~T$), symmetric otherwise ($\delta=0\degree$)\\ \hline  
    \end{tabular}
    
    \vspace{0.1cm}
    
    \begin{tabular}{|m{7em}|m{4.2em}|m {11em}|m{9em}|m {4.2em}|m{11em}|} \hline
    \multirow{4}{5em}{\makecell{Varying initial \\canting angle \\ $\theta_{m0}$, $m=6.9~\mu_B$}} & $\theta_{m0}(\degree)$ & $\Delta B/B_c=|B_c-B_{c^{\prime}}|/B_c$ &  \multirow{4}{5em}{\makecell{Varying magnitude \\of Eu moment \\$m ~(\mu_B)$, $\theta_{m0}=0\degree$}} & $m ~(\mu_B)$ & $\Delta B/B_c=|B_c-B_{c^{\prime}}|/B_c$ \\ \cline{2-3} \cline{5-6}
    & $0\degree$ &  $0.15$ &  & $6.9$ & $0.15$ \\ \cline{2-3} \cline{5-6}
    & $15\degree$ & $0.30$ &  & $3.5$ & $0.30$\\ \cline{2-3} \cline{5-6}
    & $45\degree$ & $0.57$ &  & $1.8$ & $0.50$\\ \hline
    \end{tabular}
    \label{summary}
    
\end{table*}

\vspace{0.2cm}
\noindent \textit{Results with an initial canting angle}\\
\indent In the above discussion, the initial orientation is chosen to be along the $z$ direction. Now, we consider the case when the Eu moment is canted initially at a polar angle $\theta_{m0}$. In this case, the polar canting angle varies with the in-plane field as $\theta_m = \sin^{-1}(B/B_c) + \theta_{m0}$. Fig.~\ref{fig6} shows variation in the PHC with $\theta_{m0}$. We find that increasing $\theta_{m0}$ (i) leads to a significant decrease in both the first and second critical field values and (ii) results in an increase in the difference of the critical field values. This behavior of the PHC with $\theta_{m0}$ can be understood, again by looking at the change in Berry curvature, as shown in Fig.~\ref{omega}(e). The maximum of the Berry curvature decreases and the field range within which the Berry curvature remains large increases with an increase in $\theta_{m0}$. In this case, we can express the dependence of the PHC on $\theta_{m0}$ as 
$$|B_{c^{\prime}}|\propto|B_c| \propto \frac{1}{\theta_{m0}}~~~{\rm and} ~~~|\Delta B| \propto \theta_{m0}$$

\noindent Therefore, with the variation in $m$, $\theta_m$, $\phi_m$, $\theta_{m0}$, $B_{in}$, and $\theta_{in}$, the PHC may acquire symmetric, antisymmetric, or anisotropic behavior with $\pi$ or $2\pi$ periodicity.\\

\noindent {\it Summary of the results presented in Sec. III}\\ 
\indent We discussed the anomalous planar Hall effect for restricted and unrestricted orientations of the magnetic moment due to thin film EuS in response to the external in-plane magnetic field. The first case focuses on the behavior of the planar Hall signal with different possible orientations of the easy axis alignment as summarized in panel A of Table~\ref{summary}. Our computed results show that the orientation and magnitude of the Eu moment play an important role in determining the characteristics of the PHC at the interface. In the second case, the Eu moment reorients itself with the in-plane field to minimize the free energy of the system. Our analysis follows the energy-minimized solution of canting in the presence of the external in-plane field. The polar canting angle $\theta_m$ exhibits a monotonically-increasing behavior with respect to changes in the in-plane magnetic field, and the azimuthal canting angle $\phi_m$ is considered to be (\romannumeral1\relax) fixed and (\romannumeral2\relax) changing with in-plane field angle. The dependence of the polar canting angle on the in-plane field introduces two critical fields. The first critical field is not only influenced by the second critical field but also depends on factors like the initial polar canting angle and the magnitude of the Eu moment. The second critical field is determined by the thickness of the thin film EuS, the magnitude of the Eu moment, and the initial polar canting angle. PHC is dominant for the in-plane field in the second regime, $|B_{c^{\prime}}| \leq |B_{in}| \leq |B_c|$. The nature of the PHC in the second regime is summarized in panel B of Table~\ref{summary}. 

The effect of the orientation of the Eu-moment on the periodicity of PHC, can also be understood from the perspective of symmetry breaking in the interface plane.  In the absence of any Eu moment and in the case when the Eu moment is aligned perpendicular to the interface plane, the interface plane has full $2\pi$ rotational symmetry. In the absence of any Eu moment, the PHE arises from the first term in Eq.(\ref{s_yx}) i.e. the product of the modified velocities, and the resulting PHE has $\pi$ periodicity with a maximum planar Hall conductivity near 45$\degree$ field angle. When the Eu moment is perpendicular to the interface plane, the PHE also has $\pi$ periodicity but the maximum conductivity is shifted towards higher field angles due to finite Berry curvature. In the presence of canted Eu moment in EuS, the rotational symmetry in the interface plane is broken and the peak position of the Berry curvature is shifted from the center of the Brillouin zone. Therefore, the terms in Eq.~(\ref{s_yx}) contribute differently depending on the Eu moment canting, leading to various patterns in PHC, as summarized in Table \ref{summary}.

\section{Hall effect from magnetic skyrmions at the interface}
Due to the presence of strong spin-orbit coupling in Bi$_2$Se$_3$ and broken inversion symmetry at the interface, Dzyaloshinskii-Moriya (DM)-type antisymmetric spin exchange interactions can appear on the EuS side of the interface. This DM interaction is known to produce non-trivial magnetic texture which can lead to topological Hall effect~\cite{nagaosa_NatNano2013, heinze_NatPhy2011, yokouchi_JPhysSocJpn2015, yu_Nat2010, muhlbauer_Sci2009,mohanta_PRB2019}. To explore the possibility of the formation of non-trivial spin texture at the interface and the occurrence of the topological Hall effect in our considered planar Hall geometry, we perform Monte Carlo calculations (see Appendix E for details), based on the following spin Hamiltonian

\begin{figure*}[t]
\begin{center}
\vspace{-0mm}
\epsfig{file=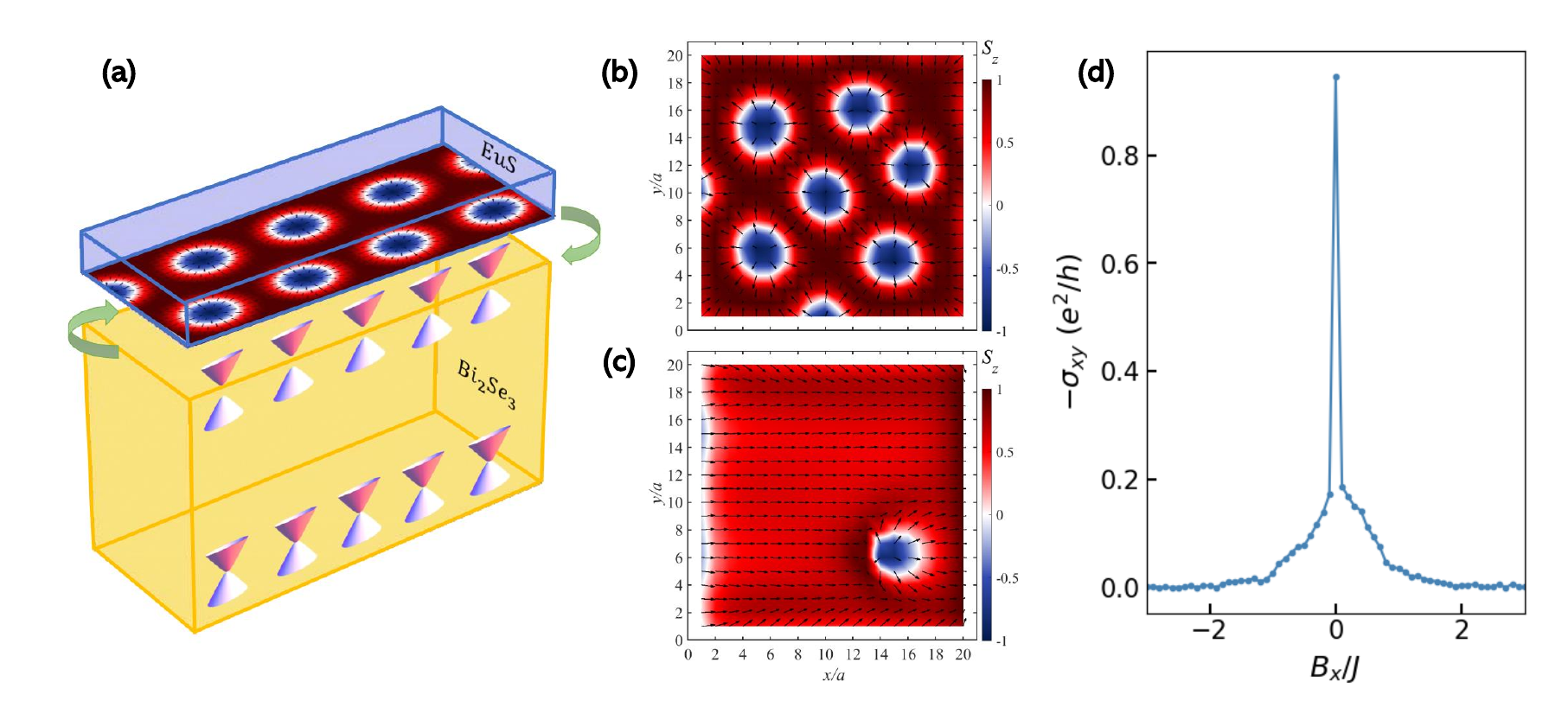,trim=0.0in 0.0in 0.0in 0.0in,clip=false, width=180mm}
\caption{(a) Illustration of the model for skyrmion formation at the $\rm{Bi}_2\rm{Se}_3/\rm{EuS}$ interface. The spin texture at the interface is depicted for the cases (b) $B_{in}=0J$ and (c) $7J$ with $J=-1$, $D=1.4J$, $m=5J$, $\theta_m=\phi_m=0\degree$, and $\theta_{in}=0\degree$. (d) Planar topological Hall conductivity plotted for $\alpha=0.5J$ and $B_{in}$ in the range $-2.5J$ to $2.5J$; other parameters are the same as above.}
\label{spin-texture}
\vspace{-4mm}
\end{center}
\end{figure*}

\begin{equation}
\begin{split}
   H_{spin} = &- J \sum_{\langle ij \rangle} {\bf S}_i \cdot {\bf S}_j - \sum_{\langle ij \rangle} {\bf D}_{ij} \cdot ({\bf S}_i \times {\bf S}_j) \\ &- \sum_{i} ({\bf m}+ {\bf B}) \cdot {\bf S}_i 
\end{split}
\end{equation}
where $J$ is the strength of ferromagnetic Heisenberg exchange coupling, ${\bf D}_{ij} = D (r_i - r_j) \times \hat{z}/|r_i - r_j|$ is the DM vector acting between $i$ and $j$ lattice sites in the $x$-$y$ plane with $D$ being the strength of DM interaction, ${\bf B}$ is the external in-plane magnetic field, and ${\bf m}$ is the magnetic moment due to EuS. Here, $i$ and $j$ represent nearest neighbor lattice sites. 

Topologically non-trivial spin textures, such as magnetic skyrmions, have been both theoretically proposed and observed in experiments via topological Hall measurements \cite{Hilary_PRB2015, Stefan_PRB2022, wu_AdvMat2020, zhang_NanoLett_2018, li_NanoLett2021}. In our considered planar Hall geometry, magnetic skyrmions can appear naturally when the Eu moment is aligned out of the interface plane, as found in the first-principle analysis of the interface~\cite{Kim_PRL2017}, or there is a suitable out-of-plane contribution to the Zeeman exchange. These magnetic skyrmions can exchange-couple to the gapped Dirac surface states on the surface of the topological insulator and produce a planar topological Hall effect. This scenario is shown schematically in Fig.~\ref{spin-texture}(a). In the absence of the applied in-plane magnetic field, a skyrmion crystal phase appears spontaneously in our Monte Carlo calculations at low temperatures, within a range of values of the out-of-plane Eu moment $m$. Fig.~\ref{spin-texture}(b) shows the spin configuration on the EuS side of the interface in the skyrmion crystal phase, obtained with parameters $J\!=\!-1$, $D\!=\!1.4J$, $m\!=\!5J$, and $B_{in}\!=\!0$. In this skyrmion crystal phase, when the in-plane magnetic field is turned on, the skyrmions get deformed due to spins canted along the field direction, and eventually beyond a critical field value an in-plane ferromagnetic phase is established. The spin texture with a deformed skyrmion at an intermediate field value is shown in Fig.~\ref{spin-texture}(c). 

To calculate the Hall conductivity arising due to the magnetic texture with a non-zero scalar spin chirality, we consider that the spin texture is coupled to the conduction electron on the surface of Bi$_2$Se$_3$. For this purpose, we consider the below Hamiltonian,
\begin{equation}
\begin{split}
     \mathcal{H}_{ex} =& -t \sum_{\langle ij \rangle} c^{\dag}_{i\sigma} c_{j\sigma} - J' \sum_{i,\sigma, \sigma'} ({\bf S}_i \cdot \bm{\sigma}_{\sigma \sigma'}) c^{\dag}_{i\sigma} c_{i\sigma'} \\&- \frac{i \alpha}{2a} \sum_{\langle ij \rangle \sigma, \sigma'} \hat{z} \cdot (\bm{\sigma} \times {\bf d}_{ij})_{\sigma \sigma'} c^{\dag}_{i\sigma} c_{j\sigma'},
\end{split}
\end{equation}
where the first term corresponds to the tight-binding hopping term, the second is the exchange coupling of the magnetic texture on the EuS side of the interface with the itinerant electrons on the Bi$_2$Se$_3$ surface, and the third term represents the Rashba hopping term with ${\bf d}_{ij}$ being the unit vector between sites $i$ and $j$, $\alpha$ the rashba coupling strength, $a$ the lattice spacing, and ${\bm \sigma} \equiv (\sigma_x,\sigma_y, \sigma_z)$ the Pauli matrices. By diagonalizing the above Hamiltonian, from the eigenvalues and eigenvectors, we calculate the transverse Hall conductivity using the Kubo formula, given by
\begin{equation}\label{kubo}
    \sigma_{xy} = \frac{e^2}{h} \frac{2\pi}{N} \sum_{\epsilon_m \neq \epsilon_n} \frac{f_m-f_n}{(\epsilon_m-\epsilon_n)^2 + \eta^2} {\rm Im} (\bra{m}\hat{j}_x\ket{n} \bra{n}\hat{j}_y\ket{m})
\end{equation}
where, $\hat{j_x}$ and $\hat{j_y}$ are the current operators in the $x$ and $y$- directions, $f_{m/n}$ is the fermi-dirac distribution at temperature $T$ and energy $\epsilon_{m/n}$, $\ket{m}$ is the $m$th eigenstate of $\mathcal{H}_{ex}$, $N(L_x \times L_y)$ is the lattice size, and $\eta$ is the relaxation rate. The transverse Hall conductivity in this case follows the familiar antisymmetric relation $\sigma_{xy} = -\sigma_{yx}$~\cite{Battilomo_PRR2021,Wang_PRL2024,Xiang_PRB2024}, which is in stark contrast with the Berry curvature -driven planar Hall effect discussed in the preceding sections. For our calculations, we have used $\alpha=0.5$, $T=0.001J$, $L_x=L_y=20$, and $\eta=0.1$.
Fig.~\ref{spin-texture}(d) shows the variation of the topological Hall conductivity with increasing amplitude of the in-plane magnetic field $B_x$, applied along the $x$ direction. The large value of $\sigma_{xy}$ at $B_x\!=\!0$ decreases monotonically with increasing $B_x$, and becomes zero at large fields.

In the above description, we have considered the scenario in which the Eu moment is aligned perpendicular to the interface plane. However, for canted Eu moment, the magnetic skyrmions may not be circular in shape, producing an anisotropy in the topological Hall response in the planar Hall geometry. This anisotropy in the spin texture breaks the rotational symmetry in the interface plane, which can lead to a $\pi$ periodicity in the topological PHE.

\section{Conclusion}
Planar Hall effect in topological insulators with a gapped surface band below the Fermi level can originate from the Berry curvature-induced anomalous velocity. On the other hand, planar topological Hall effect can arise at the interface between a topological insulator and a magnetic insulator from a spin texture with finite scalar spin chirality. In this paper, we explored both anomalous and topological Hall effects in the planar Hall geometry based on Bi$_2$Se$_3$/EuS interfaces. The canting of the Eu moment near the interface can lead to an anisotropy in the Hall conductivity with respect to the applied in-plane magnetic field. We also found that the preformed skyrmions can be distorted in the presence of the in-plane magnetic field. When the Fermi energy is just above the gapped Dirac surface band, the Berry curvature-driven Hall effect dominates. However, if the Fermi energy is situated within the bulk bands, the contribution from the non-trivial spin texture is expected to be significant, as will be presented in another study~\cite{Dhavala_2024submission}. The anisotropy in the planar Hall conductivity, when the Fermi energy is within the bulk energy gap, appears from the magnetic proximity effect of the canted Eu moment and the in-plane applied field  on the surface states of the topological insulator. Our results can be useful to determine the nature of the canting of the Eu moment using the anisotropy in planar Hall conductivity in Bi$_2$Se$_3$/EuS interfaces, and possibly in other similar systems.

\section*{Acknowledgements} 
NM acknowledges support of an initiation grant (No. IlTR/SRIC/2116/FIG) from IIT Roorkee and SRG grant (No. SRG/2023/001188) from SERB. Calculations were performed at the computing resources of PARAM Ganga at IIT Roorkee. JS was supported by Ministry of Education, Government of India via a research fellowship. KVR acknowledges support by Department of Atomic Energy, Government of India, under Project Identification No. RTI 4007, ONRG Grant No. N62909-23-1-2049, DST Grant No. DST/NM/TUE/QM-9/2019 (G), and SERB CRG Grant No. CRG/2019/003810.

\appendix
\section{Derivation of slab Hamiltonian}

The effective low energy bulk Hamiltonian for $\mathrm{Bi}_2\mathrm{Se}_3$ in the basis $\ket{P1^+_z, \uparrow}$, $\ket{P2^-_z, \uparrow}$, $\ket{P1^+_z, \downarrow}$, $\ket{P2^-_z, \downarrow}$ where $P1_z$ and $P2_z$ are $p_z$ orbitals of $\mathrm{Bi}$ and $\mathrm{Se}$, $\pm$ represents bonding and anti bonding states of orbitals is \cite{Mohanta_SciRep2017},

\begin{equation}
    H_{Bulk}(\textbf{k}) = \begin{pmatrix}
        \tilde{\epsilon}_+(\textbf{k}) & B_1 k_z & 0 & A_1 \tilde{k}_- \\ 
        B_1 k_z &\tilde{\epsilon}_-(\textbf{k}) & A_1 \tilde{k}_- & 0 \\
        0 & A_1 \tilde{k}_+ & \tilde{\epsilon}_+ (\textbf{k}) & -B_1 k_z \\
        A_1 \tilde{k}_+ & 0 & -B_1 k_z & \tilde{\epsilon}_-(\textbf{k}) \\
    \end{pmatrix}\label{bulk},
\end{equation}

\noindent where $\tilde{\epsilon}_{\pm}(\textbf{k})= \tilde{\epsilon}_0(\textbf{k}) \pm \tilde{M}(\textbf{k})$, $\tilde{\epsilon}_0(\textbf{k})=\tilde{C}_0 + \tilde{C}_1 k_z^{2} + \tilde{C}_2 (k_x^2 + k_y^2)$,  $\tilde{M}(\textbf{k})=\tilde{M}_0 + \tilde{M}_1 k_z^{2} + \tilde{M}_2 (k_x^2 + k_y^2)$, $\tilde{k}_{\pm}=k_x \pm i k_y$. 

\noindent The lattice generalized Hamiltonian is obtained by substituting $k_ia_i \simeq \sin k_ia_i$ and $(k_ia_i)^2 \simeq 2(1-\cos k_ia_i)$ in Hamiltonian given by \ref{bulk}.

\begin{align}
   H_{lattice}(\textbf{k}) = \begin{pmatrix}
        \epsilon_+(\textbf{k}) & B(k_z) & 0 & A_0 k_- \\ 
        B(k_z) &\epsilon_-(\textbf{k}) & A_0 k_- & 0 \\
        0 & A_0 k_+ & \epsilon_+ (\textbf{k}) & -B(k_z) \\
        A_0 k_+ & 0 & -B(k_z) & \epsilon_-(\textbf{k}) \\
    \end{pmatrix}\label{lattice},
\end{align}

\noindent where $\epsilon_{\pm}(\textbf{k})= \epsilon_0(\textbf{k}) \pm M(\textbf{k})$, $\epsilon_0(\textbf{k})=C_0 + 2C_1 (1-\cos k_zc) + 2C_2 (2-\cos k_xa - \cos k_ya)$,  $M(\textbf{k})=M_0 + 2M_1 (1-\cos k_zc) + 2M_2 (2-\cos k_xa - \cos k_ya)$, $k_{\pm`}=\sin k_xa \pm i \sin k_ya$, $B(k_z) = B_0 \sin k_zc$, $a$ and $c$ are lattice constants, and the modified parameters are, $A_0 = A_1/a$, $B_0 = B_1/c$, $C_0=\tilde{C}_0$, $C_1 = \tilde{C}_1/c^2$, $C_2=\tilde{C}_2/a^2$, $M_0=\tilde{M}_0$, $M_1 = \tilde{M}_1/c^2$, and $M_2=\tilde{M}_2/a^2$.

\noindent The slab Hamiltonian for $\mathrm{Bi}_2\mathrm{Se}_3$ can be obtained by partial inverse Fourier transform of lattice generalized Hamiltonian. 
\begin{equation}\begin{split}
    H_{slab}(\textbf{k})= \sum_{\textbf{k}} c_{\textbf{k}}^{\dag} H_{lattice}(\textbf{k}) c_{\textbf{k}} = \frac{1}{N^2} \sum_{\textbf{k}} \sum_{l_z, l'_z} e^{-ik_zl_z} \\c_{k_x,k_y,l_z}^{\dag} H_{lattice}(\textbf{k})  e^{ik_zl_z} c_{k_x,k_y,l_z}
\end{split}
\end{equation}

\begin{equation}\begin{split}\hspace{-1.8cm}
    H_{slab} = \frac{1}{N} \sum_{\textbf{k}} \sum_{\alpha=1}^{4} (\sum_{l_z=1}^{N_z} c_{k\alpha l_z}^{\dag} H_{0}(\textbf{k}) c_{k\alpha l_z} + \\  \sum_{l_z=1}^{N_z-1}[c_{k\alpha l_z}^{\dag} H_{1} c_{k\alpha l_z+1} + H.c.])
\end{split}
\end{equation}

\noindent The total slab Hamiltonian for $N$ quintuple layers in the basis $\ket{P1^+_z, \uparrow, l_z}$, $\ket{P2^-_z, \uparrow, l_z}$, $\ket{P1^+_z, \downarrow, l_z}$, $\ket{P2^-_z, \downarrow, l_z}$, where $l_z$ is the layer index is given by: 
\begin{equation}
    H(k) = \begin{pmatrix}
        H_0 + H_z & H_1 &  &   &   &  &  \\
        H^{\dag}_1 & H_0 & H_1 &  &  &  &  \\
          & H^{\dag}_1 & H_0 & H_1 &   &  &  \\
          & &  .&   &   &  &  \\
          &  &   &.   &  &  &  \\ 
          &  &   &   &.  &  &  \\  
          &  &  &  &   H^{\dag}_1 & H_0 & H_1\\
          &  &  & & &   H^{\dag}_1 & H_0 \\
    \end{pmatrix}\label{H_total}
\end{equation}

\noindent written in the basis $\ket{P1^+_z, \uparrow, l_z}$, $\ket{P2^-_z, \uparrow, l_z}$, $\ket{P1^+_z, \downarrow, l_z}$, $\ket{P2^-_z, \downarrow, l_z}$, where ${\bf k}_{\parallel}\! \equiv \!(k_x,k_y)$, $l_z$ is the layer index which varies between 0 and $N$. The Hamiltonians $H_0$, $H_1$ and $H_z$ describe, respectively, a single quintuple layer, coupling between two neighboring layers, and Zeeman exchange coupling on the top surface layer due to proximity effect of ferromagnetic EuS and due to the applied in-plane magnetic field. These Hamiltonians are given by

\begin{equation}
\begin{split}
&H_0  =\begin{pmatrix}
  \epsilon_0+M & 0 & 0 & A_0{\bf k_-}  \\ 
        0 & \epsilon_0-M & A_0{\bf k_-} & 0 \\ 
        0 & A_0{\bf k_+} & \epsilon_0+M & 0\\
        A_0{\bf k_+} & 0 & 0 & \epsilon_0-M \\
\end{pmatrix}, 
\label{H0}
\end{split}
\end{equation}

\begin{equation}
  H_1  =\begin{pmatrix}
  -M_1 -C_1 & i B_0/2 & 0 & 0 \\ 
        i B_0/2 & M_1 - C_1 & 0 & 0 \\ 
        0 & 0 & -M_1 - C_1 & -i B_0/2\\
        0 & 0 & -i B_0/2 & M_1 - C_1 \\
\end{pmatrix}, 
\label{H1}
\end{equation}

\begin{equation}
    H_z = \begin{pmatrix}
        h_z & 0 & h_x - ih_y & 0 \\
        0 & h_z & 0 & h_x - ih_y \\
        h_x + ih_y & 0 & -h_z & 0\\
        0 & h_x + ih_y & 0 & -h_z \\
     \end{pmatrix},  
\end{equation}
where $\epsilon_0(\mathbf{k}_{\parallel}) \!=\! C_0 + 2C_1 + 2C_2 (2-\cos(k_xa)-\cos(k_ya))$, $M_0(\mathbf{k}_{\parallel}) \!=\! M_0 + 2M_1 + 2M_2 (2-\cos(k_xa)-\cos(k_ya))$, $a$ is the lattice constant, ${\bf k_{\pm}}=\sin{(k_xa)}\pm i\sin{(k_ya)}$, $h_x \!=\! m_x + g\mu_B B_x/2$,  $h_y \!=\! m_y + g\mu_B B_y/2$, and $h_z \!=\! m_z$. $ m_x \!=\! m \sin\theta_m \cos\phi_m$, $ m_y \!=\! m \sin\theta_m \sin\phi_m$, $m_z \!=\! m \cos\theta_m$,  $B_x \!=\! B \cos\theta_{in}$, and $B_y \!=\! B \sin\theta_{in}$, ${\bf m} \! \equiv (m_x,m_y,m_z)$ represents the $\mathrm{EuS}$ magnetic moment which is assumed to be canted at a polar angle $\theta_m$ with the $z$-axis and azimuithal angle $\phi_m$ with the $x$-axis on the $x$-$y$ (interface) plane, and $B$ is the magnitude of the external in-plane magnetic field applied at an angle $\theta_{in}$ with the $x$-axis on the $x$-$y$ plane. 
The parameter values used in this paper are $A_0 = 0.8 eV, ~B_0 = 0.32 eV, ~C_0 = -0.0083 eV, ~C_1 = 0.024 eV, ~C_2 = 1.77 eV, ~M_0 = -0.28 eV, ~M_1 = 0.216 eV, ~M_2 = 2.6 eV, ~g=20,$ and $a = 4.14 {\AA}$ \cite{Mohanta_SciRep2017}.

\section{Description of canting of the Eu moment}

\noindent As discussed in the supplementary of \cite{Kim_PRL2017}, the total Hamiltonian for $\mathrm{Bi_2}\mathrm{Se_3}/\rm{EuS}$ interfaces in the presence of an external magnetic field can be written as
\begin{equation}
    H = \hbar v_f {\bm \sigma} \cdot ({\bf k} \times {\bf \hat{z}}) + {\bm \sigma} \cdot (u {\bf m}-\mu_B {\bf B}) -StM_s {\bf m}\cdot {\bf B} + H_{\mathrm{EuS}}({\bf m}) 
\end{equation}
The effective total energy   
\begin{equation}
    E_{tot}({\bf m}) = \rm{constant} - A(u {\it m_z} - \mu_B {\it B_z})^2 - S t M_s ({\bf m \cdot B}) 
    \label{etot}
\end{equation}

\noindent Considering the general configuration for the magnetic moment due to EuS, i.e., ${\bf m} \! \equiv (m_x,m_y,m_z)$ canted at a polar angle $\theta_m$ with the $z$-axis and azimuthal angle $\phi_m$ with the $x$-axis on the $x$-$y$ plane, the energy-minimized configuration with respect to polar and azimuthal canting angles is given by $\phi_m = \theta_{in}$ and $\theta_m  =\sin^{-1}(StM_sB/2Au^2m) = B/B_c, ~{\rm for}~ B< B_c ~{\rm and}~ \theta_m= 90\degree,~ {\rm for}~ B \geq B_c$. Here, $t$ is the thickness of $\rm{EuS}$, $M_s$ is saturation magnetization and $B_c$ is the critical field.

\section{Derivation of planar Hall conductivity}

\noindent The semi-classical equations of motion for a Bloch electron in the presence of electric and magnetic field are \cite{Ziman_2001, Xiao_PRL2005, Dai_PRL2017, Min_NanoLett2022, Morimoto_PRB2016, Nandy_SciRep2018}:

\begin{equation}\label{r_eq}
    \dot{{\bf r}} = \frac{1}{\hbar} \nabla_{k} \epsilon_{k} - \dot{{\bf k}} \times {\bf \Omega({\bf k})}
\end{equation}

\begin{equation}\label{k_eq}
    \hbar \dot{{\bf k}} = -e {\bf E} - e \dot{{\bf r}} \times {\bf B}
\end{equation}

\noindent After substituting equation \ref{k_eq} in equation \ref{r_eq} and conducting algebraic manipulation, the resulting equations are as follows:

\begin{equation} \label{final_r}
 \dot{{\bf r}} = D({\bf B}, {\bf \Omega(k))} ~\Big[\tilde{{\bf v}}_k + \frac{e}{\hbar} ~{\bf E} \times {\bf \Omega(k)} + \frac{e}{\hbar} ~(\tilde{{\bf v}}_k  \cdot {\bf \Omega(k)}) {\bf B}\Big] 
\end{equation}

\begin{equation}\label{final_k}
\hbar \dot{{\bf k}} = -D({\bf B}, {\bf \Omega(k)}) ~\Big[e {\bf E} + e (\tilde{{\bf v}}_k \times {\bf B} )+  \frac{e^2}{\hbar}~ ({\bf E} \cdot {\bf B})~ {\bf \Omega(k)}\Big]
\end{equation}

\noindent Where, $D^{-1}({\bf B}, {\bf \Omega(k)}) = 1 + \frac{e}{\hbar} ({\bf B} \cdot {\bf \Omega(k)})$ and
\\$\tilde{{\bf v}}_k = \frac{1}{\hbar} \nabla_k \tilde{\epsilon}_k = \frac{1}{\hbar} \nabla_k \epsilon_{k} - \frac{1}{\hbar} \nabla_k ({\bf m_{orb}} \cdot {\bf B})$ and $\tilde{\epsilon}_k = \epsilon_{k} - ({\bf m_{orb}} \cdot {\bf B}) $ are modified velocity and energy respectively.

\noindent Here, ${\bf \Omega(k)} = (\Omega_x, \Omega_y, \Omega_z)$ and ${\bf m}_{orb} = (m_{orb}^x, m_{orb}^y, m_{orb}^z)$ are the berry curvature and orbital magnetic moment (OMM) respectively given by,

\begin{equation}
    \Omega^n_{xy} = - {\rm Im.} \sum_{m \neq n} \frac{\braket{m|\partial_{k_x} H|n} \braket{n|\partial_{k_y} H|m}}{(E_m - E_n)^2}
\end{equation}
\begin{equation}
    m^n_{xy} = - \frac{e}{\hbar} {\rm Im.} \sum_{m \neq n} \frac{\braket{m|\partial_{k_x} H|n} \braket{n|\partial_{k_y} H|m}}{E_m - E_n}
\end{equation}
\\For $n^{th}$ band, $\Omega^n_i = \epsilon_{ijk} \Omega^n_{jk}$ and $m^n_i = \epsilon_{ijk} m^n_{jk}$.

\noindent Considering in-plane magnetic field with magnitude $B$ making an angle $\theta_{in}$ with the electric field applied in the $x$-direction i.e., ${\bf B}= B \cos\theta_{in}~ \hat{{\bf x}} + B \sin\theta_{in} ~\hat{{\bf y}}$ and ${\bf E}= E ~\hat{{\bf x}}$. 
Now, substituting equations \ref{final_r} and \ref{final_k} in Boltzmann transport equation, 
\begin{equation}\label{boltzmann eq}
      \dot{{\bf r}} \frac{\partial f({\bf r}, {\bf k}, t)}{\partial r} +  \dot{{\bf k}} \frac{\partial f({\bf r}, {\bf k}, t)}{\partial {\bf k}} = - \frac{f - f_0}{\tau({\bf k})} = - \frac{f_1}{\tau({\bf k})} 
\end{equation}
we get, 
\begin{equation}\label{bolt_subs}
\begin{split}
     -D({\bf B},{\bf \Omega(k)})~ [\frac{e}{\hbar} {\bf E} + \frac{e}{\hbar} (\tilde{{\bf v}}_k \times {\bf B} )+\\ \frac{e^2}{\hbar^2} ({\bf E}. {\bf B})~{\bf \Omega(k)}]~ {\bf \nabla}_k \tilde{f}_k = - \frac{\tilde{f}_k - \tilde{f}_{eq}}{\tau ({\bf k})} 
\end{split}
\end{equation}

\noindent Keeping only linear terms in the applied field and using the ansatz $\tilde{f_k} = \tilde{f_{eq}} + e E D \tau ~\big[\tilde{v}_x + \frac{e B \cos \theta_{in}}{\hbar} ( \tilde{{\bf v}}_k \cdot {\bf \Omega(k)})\big]~ \partial_{\tilde{\epsilon}}\tilde{f}_{eq}$ we arrive at the following equation, 

\begin{equation}
\begin{split}
    De ~\Big[ E v_x 
    + \frac{e}{\hbar} ({\bf E} \cdot {\bf B}) ~( \tilde{{\bf v}}_k \cdot {\bf \Omega(k)}) ~\partial_{\tilde{\epsilon}}\tilde{f}_{eq} 
    + \\ 
     \frac{B}{\hbar}( - \tilde{v}_z \sin \theta_{in} \frac{\partial}{\partial k_x}  
    + \tilde{v}_z \cos \theta_{in} \frac{\partial}{\partial k_y} 
    + \\ (\tilde{v}_x \sin \theta_{in} - \tilde{v}_y \cos \theta_{in} ) \frac{\partial}{\partial k_z}) ~ \tilde{f}_k \Big]
    = - \frac{\tilde{f}_{eq} - \tilde{f}_k}{\tau({\bf k})}
\end{split}
\end{equation}

\noindent Where $\tilde{f}_{eq}$ is equilibrium Fermi-Dirac distribution with modified energy. The charge density and current density modified to preserve Liouville's theorem are given by \cite{Xiao_PRL2005}:

\begin{equation}\label{rho}
    \rho = - \frac{e}{(2 \pi)^d} \int d{\bf k} D^{-1} \tilde{f}_k 
\end{equation}

\begin{equation} \label{j}
    J = - \frac{e}{(2 \pi)^d} \int d{\bf k} (D^{-1} \dot{{\bf r}} + {\bf \nabla_r} \times {\bf m_k}) \tilde{f_k} 
\end{equation}

\noindent Where, $d$ is the dimension. Using the expression $J_a = \sigma_{ab} E_b$ and using the ansatz for $\tilde{f}_k$, we get longitudinal and transverse PHC expressions as, 

\begin{equation}
\begin{split}
    \sigma_{xx} &= - \frac{e^2 }{(2 \pi)^d} \int d{\bf k} ~ D \tau \Big[\tilde{v}_x \\ &+  \frac{e B \cos \theta_{in}}{\hbar} (\bf{\tilde{ v}_k}  \cdot \Omega(\bf{k}))\Big]^2 ~\partial_{\tilde{\epsilon}} \tilde{f}_{eq}, 
\end{split}
\end{equation}

\begin{equation}
\begin{split}
\sigma_{xy} =& - \frac{e^2}{(2 \pi)^d} \int d{\bf k} ~ D \tau ~\Big[\tilde{v}_y \tilde{v}_x + \frac{e B \cos \theta_{in} ~ \tilde{v}_y }{\hbar} (\bf{\tilde {v}_k} \cdot \Omega(\bf{k})) \\ 
&+\frac{eB \sin \theta_{in} ~ \tilde{v}_x }{\hbar} ({\bf \tilde{v}_k}  \cdot {\bf \Omega(k)}) \\
&+\frac{e^2 B^2 \cos \theta_{in} \sin \theta_{in}}{\hbar^2} (\bf{ \tilde{v}_k} \cdot \bf{ \Omega( k)})^2\Big] \partial_{\tilde{\epsilon}} \tilde{f}_{eq},
\end{split}
\end{equation}  

\section{Calculation of Berry curvature components $\Omega_x$ and $\Omega_y$} 

The total slab Hamiltonian is $k_z$ independent, therefore $\Omega_x$ and $\Omega_y$ are calculated using below approximation, treating the layer index $l_z$ as the third variable:
$$\frac{\partial}{\partial k_z} \simeq \frac{\partial}{\partial \frac{1}{l_z}} = -l_z^2 \frac{\partial}{\partial l_z}$$ 

\noindent $\Omega_x$ and $\Omega_y$ for perpendicular Zeeman coupling, gives Berry curvature dipole centered at the gamma point in the $k$-space. The in-plane components $\pm h_y (h_x)$  shifts the centre of ${\bf \Omega}{\bf(k)}$ to $\pm \delta{k_y} (- \delta{k_x})$.

\vspace{0.1cm}
\section{Monte Carlo annealing method}
Low temperature spin configurations are obtained using Monte Carlo (MC) annealing calculations on a square lattice of size 20$\times$20 with open boundary conditions. The annealing procedure was started at a high temperature $T\!=\!5J$ with an initial random spin configuration, and the system was gradually cooled down from $T\!=\!5J$ to $T\!=\!0.001J$ in 100 steps. At each temperature step, $10^8$ MC spin update steps are performed. At each spin update step, a new orientation of the spin vector at a randomly-selected site is chosen randomly within a small cone of angle $\Delta \theta=2^{\circ}$ around the initial spin direction. The new {\it trial} spin configuration is accepted or rejected using the standard metropolis algorithm: accepted if the energy difference $\Delta E$ between the two spin configurations is negative, and accepted with the Boltzmann probability $e^{-\Delta E/(k_BT)}$ when $\Delta E$ is positive; the latter is to incorporate the thermal spin fluctuation which increases with temperature. Once the spin configuration is stabilized (\textit{e.g.} in the skyrmion crystal phase) at the lowest temperature $T=0.001J$, a magnetic field along the in-plane $x$ direction is applied, and $10^{10}$ MC spin updates are conducted at various values of the in-plane field amplitude $B_x$ to obtain the spin configuration. 

\vspace{0.3cm}


%

\end{document}